\documentclass{aastex631}

\usepackage{tabularx}
\usepackage{xspace}
\usepackage{array}
\usepackage{nameref}
\usepackage[utf8]{inputenc}
\usepackage[T1]{fontenc}
\usepackage{babel}
\usepackage{amsmath}
\usepackage{amssymb}
\usepackage{graphicx}
\usepackage{url}
\usepackage{multirow}
\usepackage{hyperref} 
\usepackage{orcidlink}
\usepackage{footmisc}

\newcommand{\GEO}{GEO\,600}

\begin{document}

\title{Open data from the third observing run of LIGO, Virgo, KAGRA and GEO}



\author{R.~Abbott}
\affiliation{LIGO Laboratory, California Institute of Technology, Pasadena, CA 91125, USA}
\author{H.~Abe}
\affiliation{Graduate School of Science, Tokyo Institute of Technology, 2-12-1 Ookayama, Meguro-ku, Tokyo 152-8551, Japan  }
\author{F.~Acernese}
\affiliation{Dipartimento di Farmacia, Universit\`a di Salerno, I-84084 Fisciano, Salerno, Italy}
\affiliation{INFN, Sezione di Napoli, I-80126 Napoli, Italy}
\author[0000-0002-8648-0767]{K.~Ackley}
\affiliation{OzGrav, School of Physics \& Astronomy, Monash University, Clayton 3800, Victoria, Australia}
\author{S.~Adhicary}
\affiliation{The Pennsylvania State University, University Park, PA 16802, USA}
\author[0000-0002-4559-8427]{N.~Adhikari}
\affiliation{University of Wisconsin-Milwaukee, Milwaukee, WI 53201, USA}
\author[0000-0002-5731-5076]{R.~X.~Adhikari}
\affiliation{LIGO Laboratory, California Institute of Technology, Pasadena, CA 91125, USA}
\author{V.~K.~Adkins}
\affiliation{Louisiana State University, Baton Rouge, LA 70803, USA}
\author{V.~B.~Adya}
\affiliation{OzGrav, Australian National University, Canberra, Australian Capital Territory 0200, Australia}
\author{C.~Affeldt}
\affiliation{Max Planck Institute for Gravitational Physics (Albert Einstein Institute), D-30167 Hannover, Germany}
\affiliation{Leibniz Universit\"{a}t Hannover, D-30167 Hannover, Germany}
% ...

\collaboration{3000}{The LIGO Scientific Collaboration, the Virgo Collaboration, and the KAGRA Collaboration}



\newcommand{\GSTLAL}{GstLAL\xspace}
\newcommand{\CWB}{cWB\xspace}
\newcommand{\PYCBC}{PyCBC\xspace}
\newcommand{\MBTA}{MBTA\xspace}
\newcommand{\SPIIR}{SPIIR\xspace}
\newcommand{\MBTAONLINE}{MBTAOnline\xspace}
\newcommand{\LALSUITE}{LALSuite\xspace}
\newcommand{\LALINFERENCE}{LALInference\xspace}
\newcommand{\IDQ}{iDQ\xspace}
\newcommand{\BAYESWAVE}{BayesWave\xspace}
\newcommand{\BAYESTAR}{Bayestar\xspace}
\newcommand{\RIFT}{RIFT\xspace}
\newcommand{\PESUMMARY}{PESummary\xspace}
\newcommand{\GWCELERY}{GWCelery\xspace}

\newcommand{\BILBY}{{Bilby}\xspace}
\newcommand{\DYNESTY}{{Dynesty}\xspace}
\newcommand{\BILBYPIPE}{{BilbyPipe}\xspace}
\newcommand{\PBILBY}{{Parallel \BILBY{}}\xspace}
\newcommand{\ASIMOV}{{Asimov}\xspace}
\newcommand{\GWSUBTRACT}{{gwsubtract}\xspace}

\newcommand{\DMT}{{DMT}\xspace}
\newcommand{\DQR}{{DQR}\xspace}
\newcommand{\DQSEGDB}{{DQSEGDB}\xspace}
\newcommand{\GWDETCHAR}{{gwdetchar}\xspace}
\newcommand{\HVETO}{{hveto}\xspace}
\newcommand{\OMICRONSCAN}{{Omicron}\xspace}

\newcommand{\HDF}{{HDF5}\xspace}
\newcommand{\PYTHON}{{Python}\xspace}
\newcommand{\NUMPY}{{NumPy}\xspace}
\newcommand{\SCIPY}{{SciPy}\xspace}
\newcommand{\PLT}{{Matplotlib}\xspace}
\newcommand{\SEABORN}{{seaborn}\xspace}
\newcommand{\GWPY}{{GWpy}\xspace}
\newcommand{\GWSUMM}{{GWSumm}\xspace}

\begin{abstract}
The global network of gravitational-wave observatories now includes
five detectors, namely LIGO Hanford, LIGO Livingston,
Virgo, KAGRA, and \GEO. These detectors
collected data during their third observing run, O3, composed of three phases:
O3a starting in April of 2019 and lasting six months,
O3b starting in November of 2019 and lasting five months,
and O3GK starting in April of 2020 and lasting 2 weeks. In this paper we
describe these data and various
other science products that can be freely accessed through the
Gravitational Wave Open Science Center at \url{https://gwosc.org}.
The main dataset, consisting of the gravitational-wave strain time series
that contains the astrophysical signals, is released
together with supporting data useful for their analysis
and documentation, tutorials, as well as analysis software packages.
\end{abstract}

\section{Introduction}

Gravitational-wave (GW) detectors develop through successive generations of instruments with increasing sensitivity \citep{abbott20:_prosp_advan_ligo_advan_virgo_kagra}.
The US-based Advanced LIGO\footnote{Laser Interferometer Gravitational-Wave
Observatory} detectors \citep{aasi15:_advan_ligo} were the first
two instruments of the current generation to begin operation,
collecting data during the first observing run
(O1) from September 2015 to January 2016, including the first
direct detection of gravitational waves \citep{abbott16:_obser}.
The second observing run (O2) followed from November 2016 to August
2017, with the European detector Advanced Virgo
\citep{acernese15:_advan_virgo} joining in August 2017.
The \GEO~detector in Germany \citep{Dooley:2015fpa} serves as a center of research and development, and is used to test a number of critical detector
technologies. Another GW detector, the Japan-based
KAGRA \citep{Akutsu:2021KAGRA}, has
also been rapidly developing.

This article focuses on the data collected during 
the third observing run, O3, that took place from
April 1 2019 to April 21 2020. The bulk of this observing run
collected data only from LIGO and Virgo, and 
is divided into two main operational phases:
O3a from April 1 2019
to October 1 2019, and O3b from November 1 2019 to March 27 2020,
with a one-month
maintenance break between the two phases. KAGRA 
was expected to
join O3, but this initial plan changed due to the outbreak of COVID-19.
Instead, KAGRA and \GEO~operated during an
extended observing phase, O3GK, from April 7 to April 21 2020
\citep{kagra-GEO600-2022}.

The analysis of the O3 data has led to numerous publications. 
Those include several updates to the GWTC 
\citep[GW Transient Catalog;][]{2021PhRvX..11b1053A,LIGOScientific:2021usb,2021arXiv211103606T}
that compiles transient sources analyzed and reported by the combined LIGO-Virgo-KAGRA
Collaboration (LVK). The cumulative GWTC 
catalog currently includes nearly 100 candidate sources (with a probability of astrophysical origin > 50 \%), all associated with
the coalescence of compact star binaries composed of either neutron stars,
black holes, or both.

Following the policy defined in the LIGO Data Management Plan \citep{LIGO-DMPv2}
and a Memorandum of Understanding \citep{LIGOVirgo-MOU},
the O3 data set and 
associated science products are published through the Gravitational-Wave
Open Science Center (GWOSC) at \url{https://gwosc.org}
\footnote{This website is also accessible at \url{https://gw-openscience.org}.}
allowing the reproducibility of the analyses performed by the LVK and increasing
the impact of the data through its wider use.
This paper provides a description of the publicly released data
\citep{gwosc-O3a,gwosc-O3b,gwosc-O3GK} along with additional information on their usage.

To date hundreds of scientific articles have been written using the
data available from the GWOSC website (all datasets combined) \footnote{See \url{https://gwosc.org/projects} for a list of articles that refer to the data published on the GWOSC website.}. These analyses confirm,
complement, and extend the results published by the LVK Collaboration, demonstrating the impact on the scientific
community of the GW data releases.

This paper is organized as follows. Section ~\ref{sec:methods} 
summarizes the status of the detectors during the observing run O3, together
with high-level indicators such as their distance reach and duty cycle
of operation. This section also provides insights about how the data
are collected and calibrated, about data quality and about simulated signal
injections. Section ~\ref{sec:record} describes the format, content and
provenance of the strain data files distributed through the GWOSC, including the
nomenclature used for the calibration versions and channel
names. Section ~\ref{sec:eventportal} describes the Event Portal,
a searchable GW
event database accessible online. Details about the technical validation and review of the data and documentation are given in Section ~\ref{sec:validation}.  Finally,
Section \ref{sec:usage} provides some guiding principles to the novice user
and suggests software tools that can be used to analyze the data.


\section{Instruments}
\label{sec:methods}

The Advanced LIGO \citep{aasi15:_advan_ligo} and Advanced Virgo
\citep{acernese15:_advan_virgo} detectors are enhanced Michelson
interferometers with arm lengths of 4 km and 3 km, respectively.
Advanced LIGO comprises two detectors located at
two different sites in the US, namely, in Hanford, WA and Livingston LA,
while Advanced Virgo has a single site in Cascina, close to Pisa, Italy.
The various instrument upgrades realized between the
science runs O2 and O3 for the LIGO and Virgo detectors are described
in \citep{PhysRevD.102.062003,2021PhRvX..11b1053A,2021arXiv211103606T, Virgo:2022fxr}.
They involve many parts of the instruments, including the main laser
source and the core optics along with the installation of mitigation systems
for a range of technical noises.  One of the major
novelties in O3 both for LIGO and Virgo is the use of squeezed light
sources (see \citet{PhysRevLett.123.231107} for LIGO and
\citet{PhysRevLett.123.231108} for Virgo), a technique
\citep{squeezing1, squeezing2} that significantly reduces quantum noise,
thus enhancing the sensitivity at high frequency.

\GEO~\citep{Dooley:2015fpa}
is a British–German interferometric GW detector with 600~m arms
located near Hannover, Germany. As in LIGO and Virgo, quantum squeezing 
is used to reduce noise in the output measurement quadrature \citep{Lough:2020xft}.
This technique was first demonstrated by \GEO~\citep{LIGOScientific:2011imx}.
KAGRA is an underground laser interferometer with 3-km arms, located
underground at the Kamioka Observatory in Gifu Prefecture, Japan. 
An important feature of its design is the cooling system intended to
bring the large mirrors of the interferometer to cryogenic temperature (around 20 K) in order to reduce thermal noise~\citep{Akutsu:16,Chen:2014}. During the O3GK run however, the detector was operated at room temperature
\citep{Akutsu:2018KAGRA,Akutsu:2021KAGRA}. 

\subsection{Detectors performance}
A GW detector's performance is often globally characterized by two
measures: its duty factor, defined as the fraction of time the
detector is recording observational quality data,
and its distance reach, conventionally
measured as the binary neutron star (BNS) inspiral range
\citep{PhysRevD.47.2198,QCG38055010}, the distance to which
a BNS inspiral could be detected
with signal-to-noise ratio of 8, assuming 1.4 solar mass
component objects and averaging over source position and orientation.
The choice of this metric is a standard convention. 
The value of 1.4 solar masses is close to the measured masses of the stars in the Hulse--Taylor binary \citep{Weisberg:2016jye} and within the narrow range predicted by stellar evolution for neutron-star masses.
The distance reach of the detectors strongly depends on the source mass. 
For example, binary black-hole (BBH) systems can typically be detected 
at much greater distances, up to several Gpc \citep[e.g.,][Table~IV]{2021arXiv211103606T}.

\begin{figure}
  \includegraphics[width=0.32\textwidth]{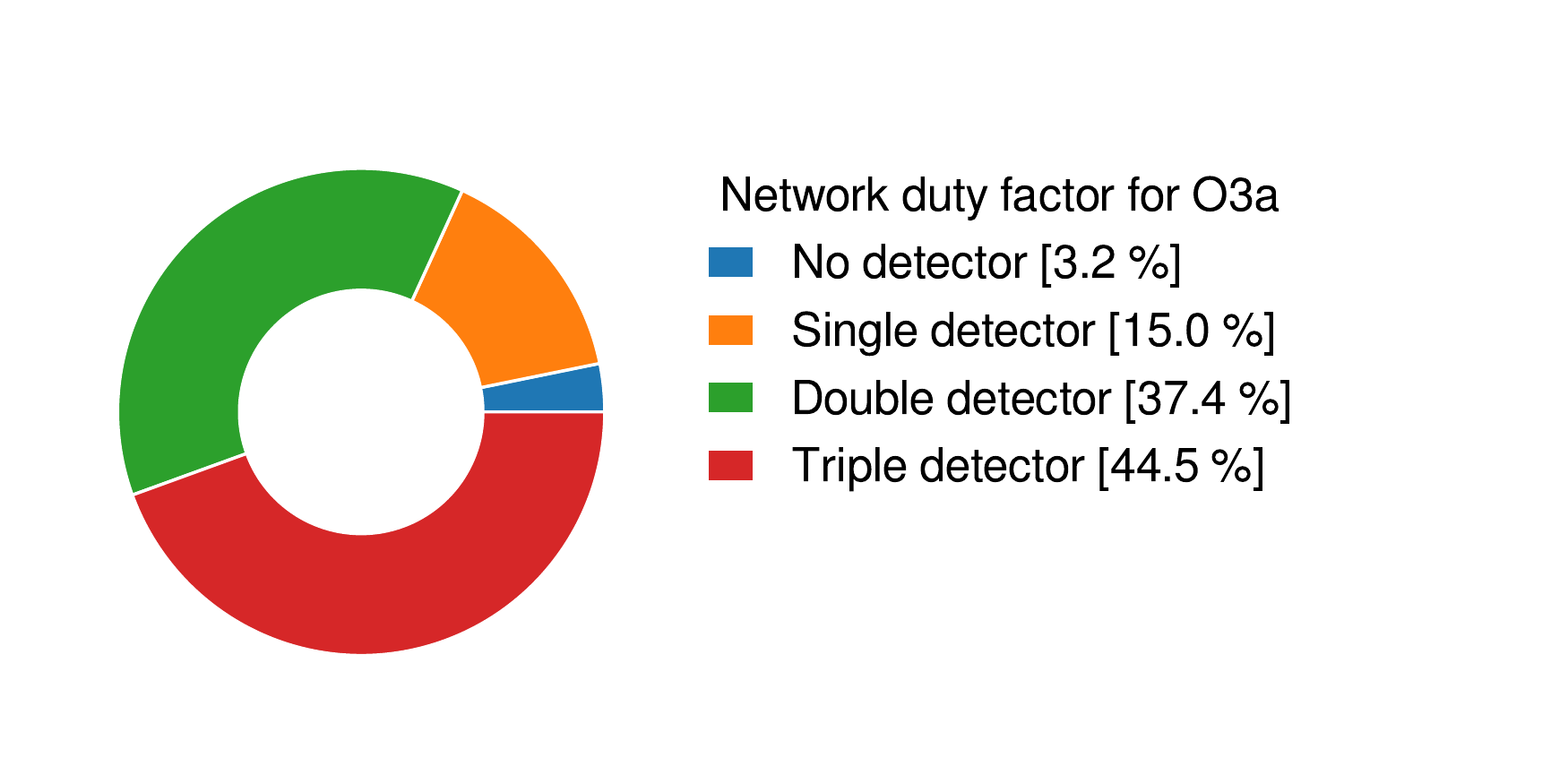}
  \includegraphics[width=0.32\textwidth]{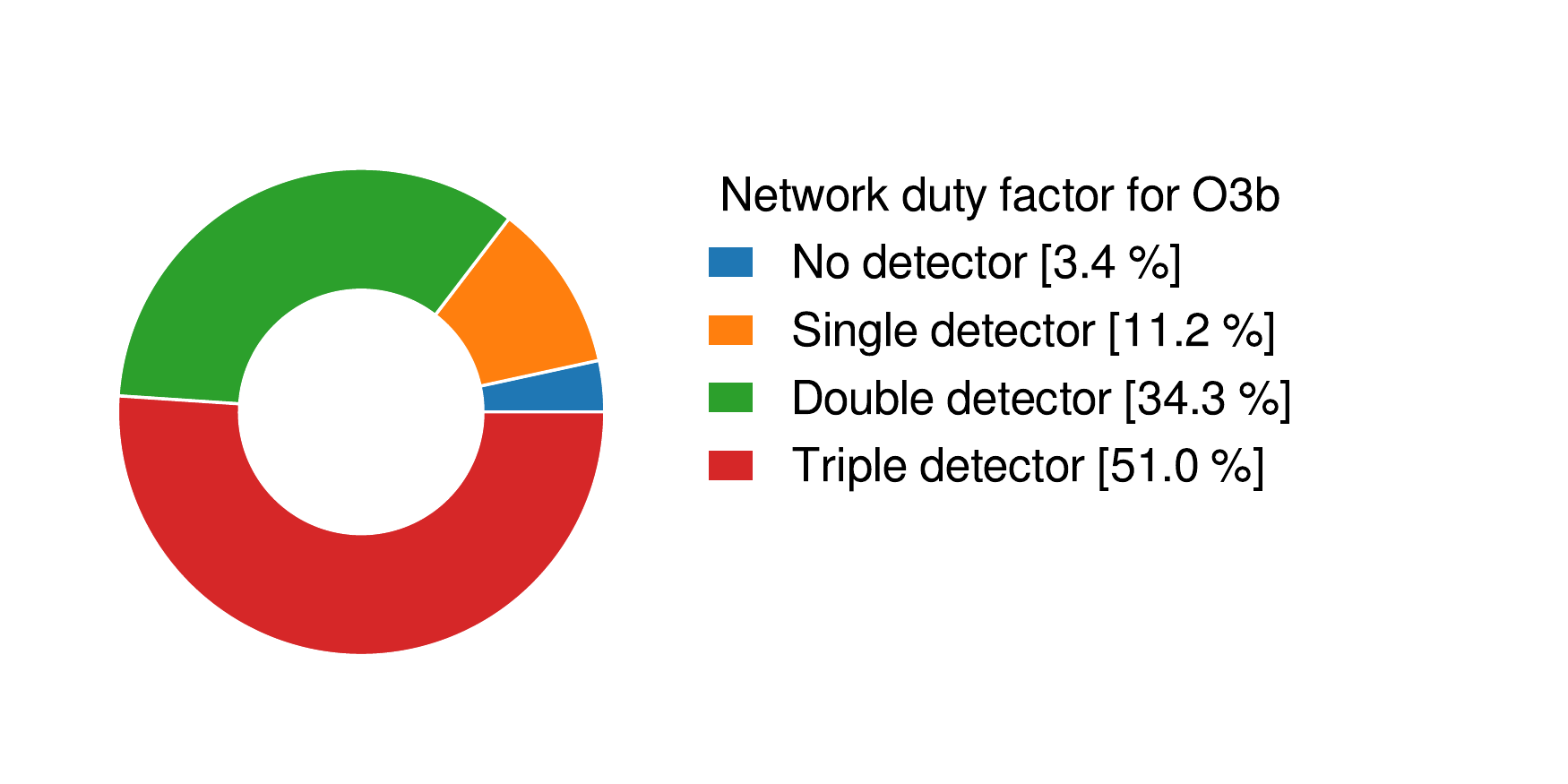}
  \includegraphics[width=0.32\textwidth]{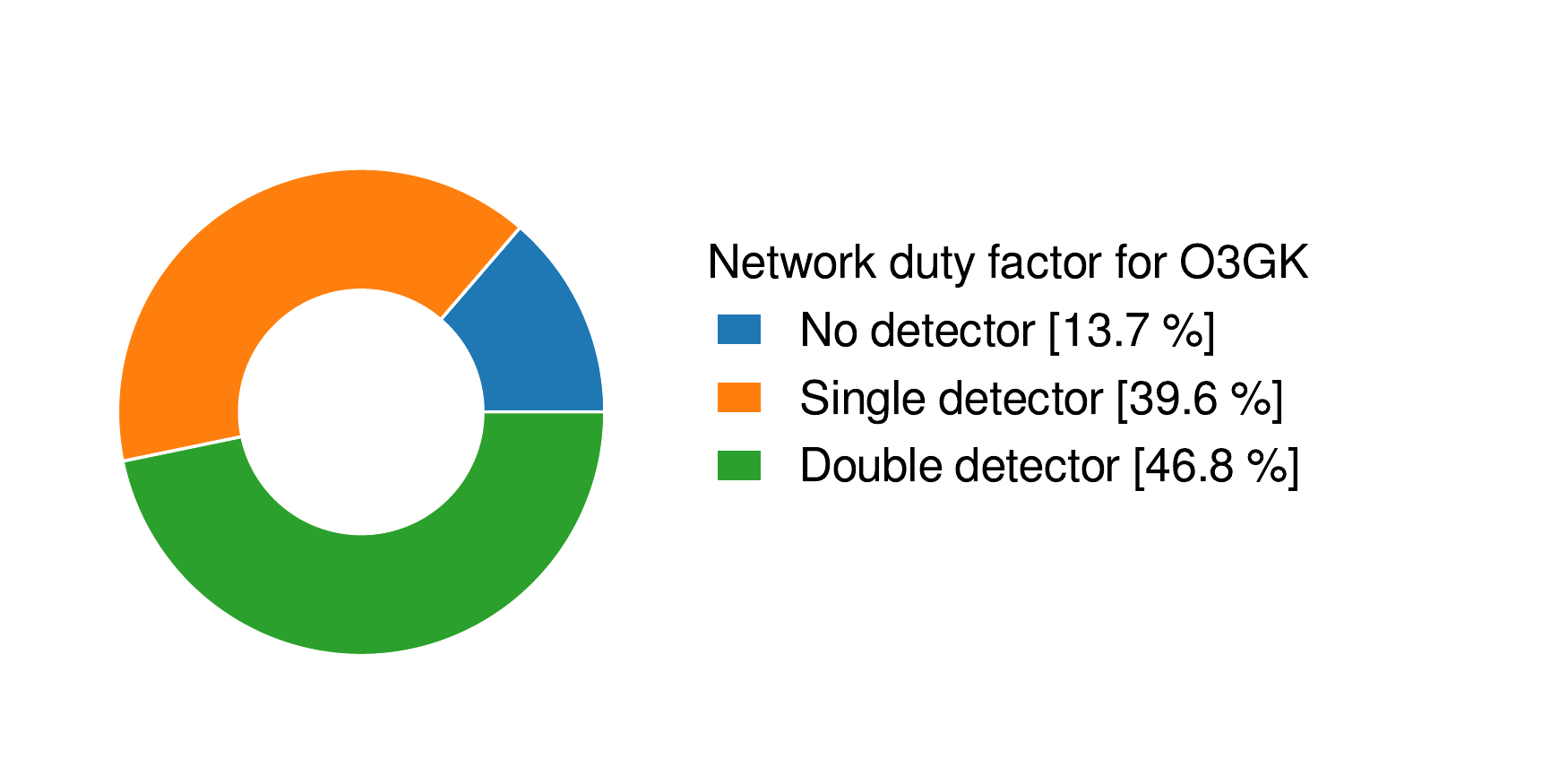}
\centering
\caption{Duty factors for the LIGO and Virgo detector network during O3a (left) and O3b (center) and for the KAGRA and \GEO~detector network during O3GK (right). These factors measure the fraction of time spent as a function of the number of detectors in operation. The same plots (caveat a difference in the color code) can be found on the GWOSC web summary pages for O3a\footref{fn:summary_o3a} and O3b\footref{fn:summary_o3b} and have been produced from \citet{kagra-GEO600-2022} for O3GK.}
\label{fig:dutyfactor}
\end{figure}

The GWOSC website hosts summary pages for
O3a\footnote{\label{fn:summary_o3a}\url{https://gwosc.org/detector_status/O3a}}
and
O3b\footnote{\label{fn:summary_o3b}\url{https://gwosc.org/detector_status/O3b}}
which describe the LIGO and Virgo operations and sensitivity. The duty
factors during O3a are 71\% for LIGO Hanford (H1), 76\% for LIGO Livingston (L1)
and 76\% for Virgo (V1). During O3b, the corresponding percentages are
79\%, 79\% and 76\%, respectively. Those translate into the observing
factors shown in Fig.~\ref{fig:dutyfactor} that quantify the
fraction of observing time spent with one, two or three instruments in
operation.

During the O3GK run, the duty factors of KAGRA (K1) and \GEO~(G1) are 53\% and
80\% respectively, leading to a coincident observing factor of
47\% \citep{kagra-GEO600-2022}. The lower duty cycle of KAGRA is
due to the fact that alignment sensing and control with wavefront
sensors was not yet implemented at the time of the run,
leading to a higher susceptibility to microseismic ground
vibrations.

\begin{figure}
  \includegraphics[width=0.8\textwidth]{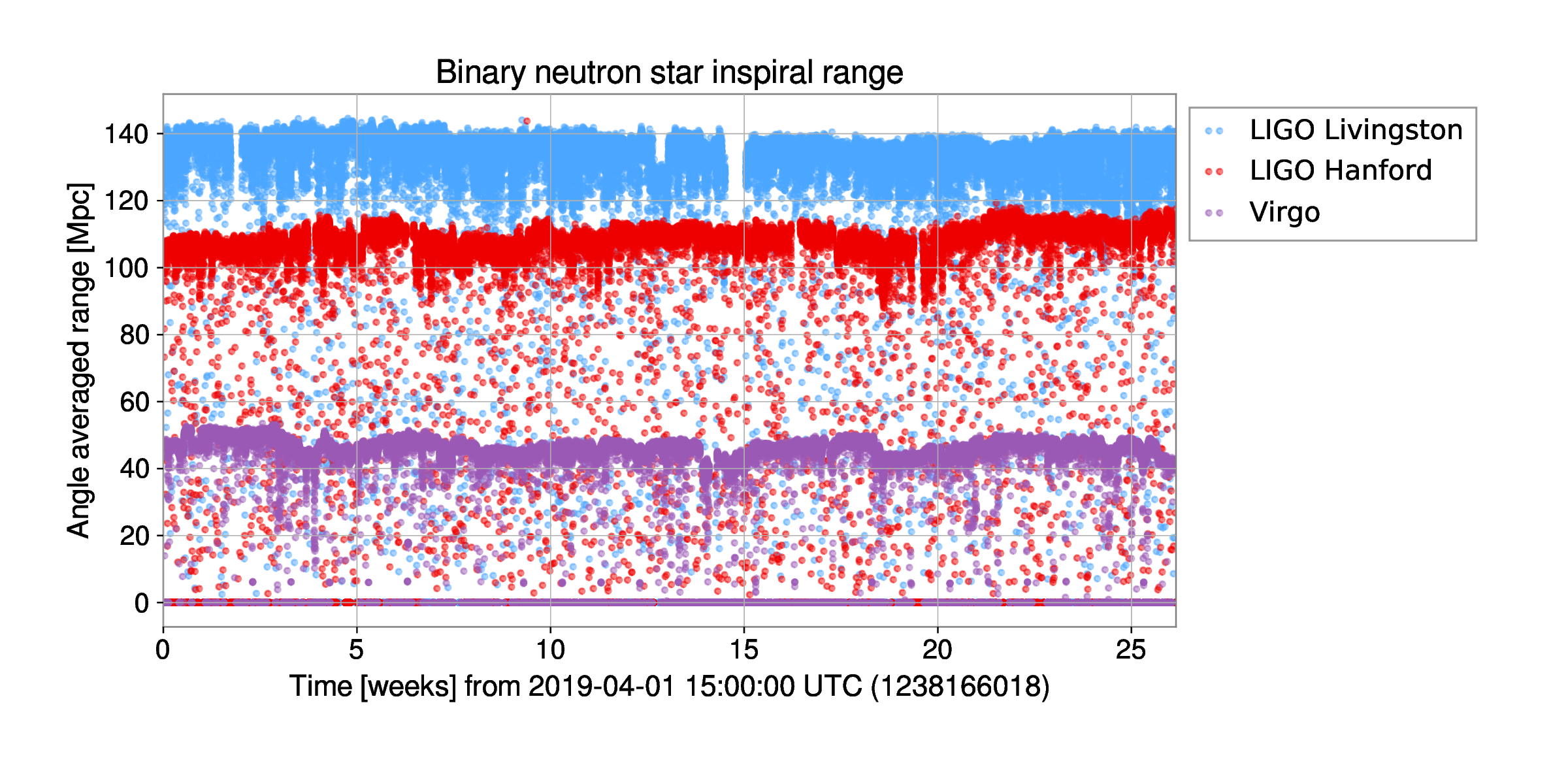}
  \includegraphics[width=0.8\textwidth]{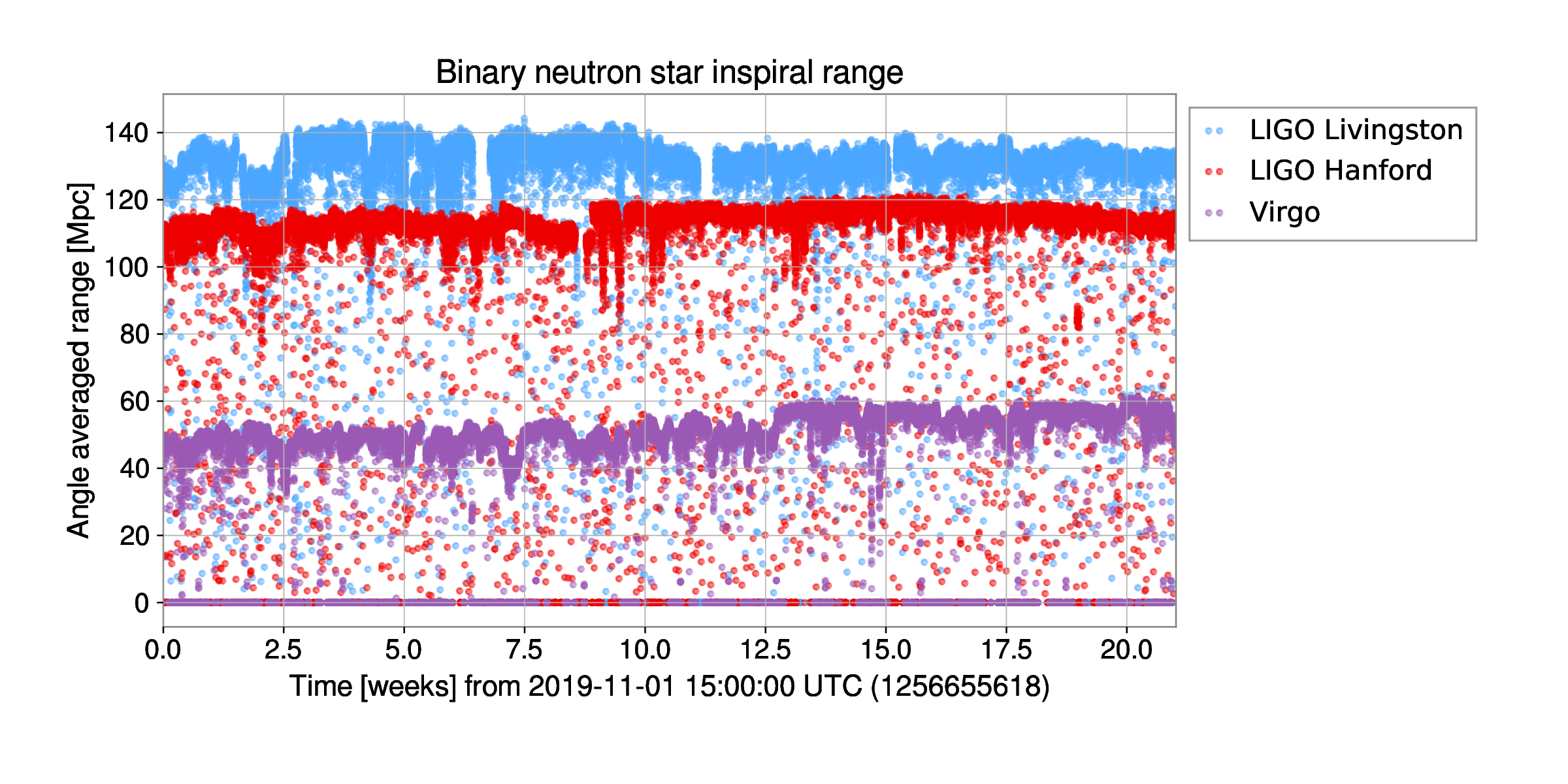}
  \includegraphics[width=0.8\textwidth]{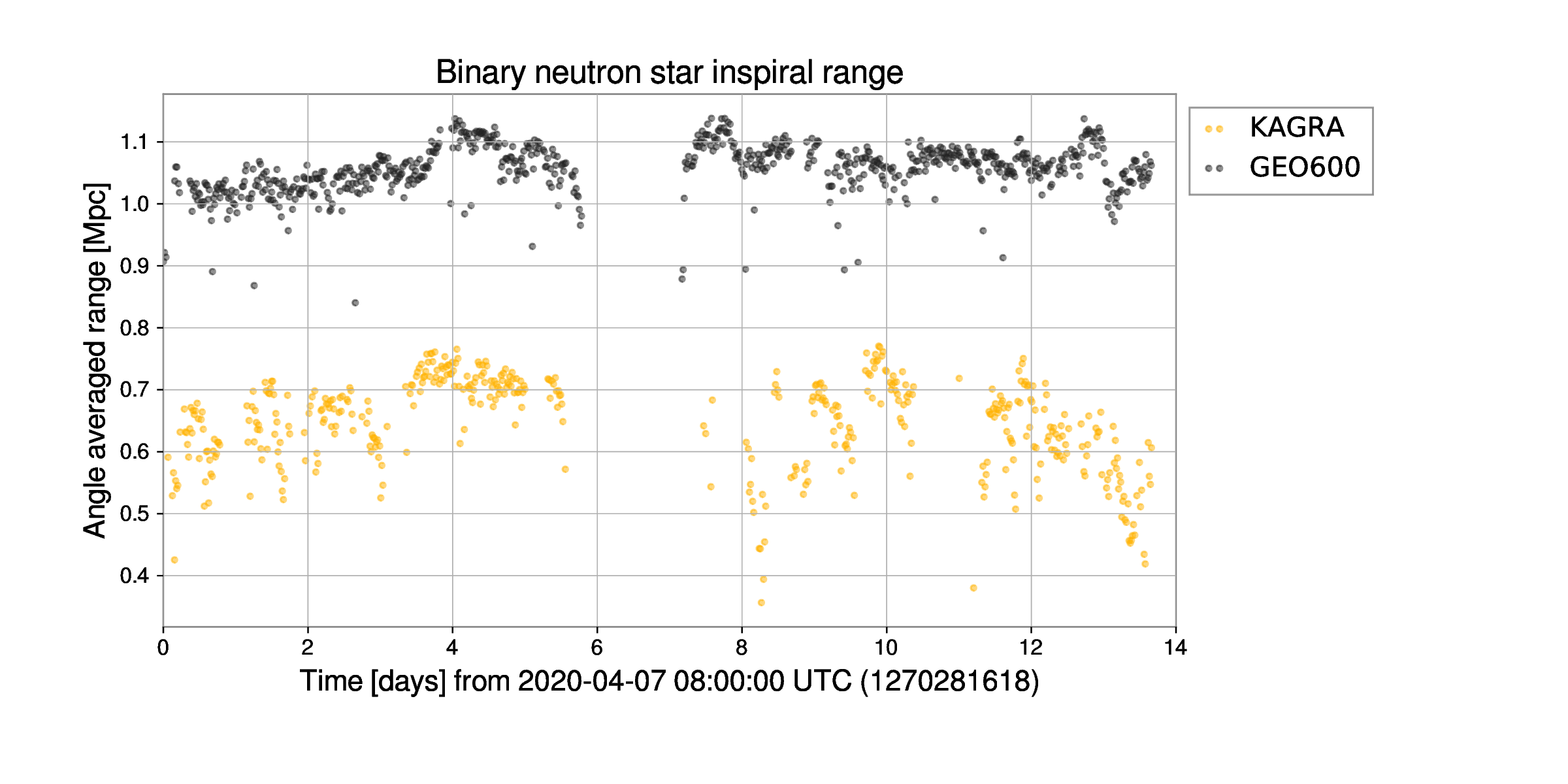}  
\centering
\caption{Binary neutron star ranges for O3a (upper plot), O3b (middle plot) with LIGO Hanford (red), LIGO Livingston (blue) and Virgo (purple) and for O3GK (bottom plot) with \GEO~(black) and KAGRA (yellow). Similar plots (besides style differences) can be found on the GWOSC web pages for O3a \footref{fn:summary_o3a} and O3b \footref{fn:summary_o3b} and in \citet{kagra-GEO600-2022} for O3GK.}
\label{fig:bnsrange}
\end{figure}

The median values of the BNS range over the whole observing run are
108 Mpc, 135 Mpc and 45 Mpc for H1, L1 and
V1 respectively during O3a, and 115 Mpc, 133 Mpc and 51 Mpc during O3b
for the same detectors. The median values of the BNS range over the O3GK period are 0.66~Mpc
for KAGRA and 1.06~Mpc for \GEO. Fig.~\ref{fig:bnsrange}
displays the median BNS range computed over regular intervals (5-minute
scale for LIGO and Virgo and 20-minute scale for \GEO~and KAGRA). The
drops that can be observed in both plots are due to transient noise
artifacts (discussed in Sec.~\ref{sec:detchar}) reducing the detector
sensitivity temporarily.  The BNS range shown in the recent GWTC
publications such as \citet{2021PhRvX..11b1053A} (Fig.~3) and
\citet{2021arXiv211103606T} (Fig.~3) are averaged over a longer period
(1 hour) and are thus less affected by transient noise.  The
longer gaps in the BNS inspiral range are due to maintenance
intervals, instrumental issues, and earthquakes.

\subsection{Calibration} 
\label{sec:calibration}

The GW strain $h(t)$ is obtained and calibrated from
variations of the optical power measured at the output port of each
detector. The calibration procedure and the corresponding
characterization of the systematic and statistical uncertainties are
described in \citet{Viets:2017yvy, Sun_2020, Sun_2021} for Advanced LIGO and
\citet{Acernese_2022} for Advanced Virgo.
Calibration is performed in two stages: an initial, online
calibration used for low-latency analysis, and a final, offline
calibration that applies any needed corrections to the initial result.
The offline calibration may correct for computer failures,
incomplete modelling of the detectors, or any systematic
errors characterized after the observing period.  The uncertainties in
the calibration procedure for both the magnitude
and phase of $h(t)$ as a function of frequency are documented \citep{calib_uncertainty_O1-O3}.

The calibration process also includes a noise subtraction step based on
independent measurements of a range of noise sources by witness
sensors, as described in \citet{Davis_2019, PhysRevD.101.042003,
  PhysRevD.101.102006, Virgo_calib}; and \citet{Acernese_2022}.
For the last two weeks of O3a, the Virgo data were reprocessed with a
new configuration of the noise subtraction \citep{Virgo_calib_new,
  Acernese_2022} so a different calibration is available just for this
period (see Table~\ref{tab:source}).

GWOSC releases two types of strain data: bulk data
spanning an entire observing run, and smaller data snippets
around the time of each GW event.
Data snippets are based on the calibration version
available at the time of publication of the related GW event.
Events that appear in multiple publications
may have multiple data snippets available, sometimes with
different calibration versions. 
Naturally, the time segments
released as data snippets are also available in the bulk data set,
but the bulk data of the entire O3a, O3b and O3GK
observation runs
provided through GWOSC correspond to the final (most up-to-date)
calibration. 
These differences in calibration can lead to discrepancies between the data snippets and the corresponding data in the bulk data release, potentially leading in turn to differences in the source parameter values that can be estimated from the data. However, as discussed in Sec.~\ref{sec:AlternateStrain}, in addition to the main bulk data release, several alternate strain channels with different calibration versions are also made public.

The detector strain $h(t)$ in O3 is calibrated only between 10 Hz and
5000 Hz for Advanced LIGO, between 20 Hz and 2000 Hz for Advanced
Virgo, between 30 Hz and 1500 Hz for KAGRA and between 40 Hz and 6000
Hz for \GEO. Any apparent signal outside these ranges cannot be
trusted because it is not a faithful representation of the GW strain
at these frequencies.
In addition, Advanced Virgo data between 49.5 Hz and 50.5 Hz are characterised by a large increase of calibration errors because of effects related to the mains power lines \citep{Acernese_2022}.
Because of this increased systematic error, data in this narrow frequency band were considered to be uninformative for source-parameter estimation (see Appendix E of \citet{2021arXiv211103606T} for relevant methods).

\subsection{Detector noise characterization and data quality}
\label{sec:detchar}

The data are dominated by instrumental noise that can be well
described as Gaussian and stationary over limited time scales and
frequency ranges. The data also contain intermittent
short-duration noise artifacts, or glitches, that contribute to the
noise background as well.  Any analysis of GW data must account
for the presence of these various noise components (see
Sec.~\ref{sec:usage} for more information about using the data).
A summary
of efforts to characterize data quality in O3 can be seen in
\citet{LIGO:detcharO2O3} for Advanced LIGO,
\citet{Virgo:2022fxr} for Advanced Virgo,
and \citet{KAGRAnoise} for KAGRA.  The overall
quality of data for transient searches is recorded
as data quality segments, described in more detail in
Sec.~\ref{sec:DqInj}.

\subsection{Signal injections}

Hardware injections are simulated GW signals added by
physically displacing the test masses (i.e. the interferometer
mirrors) \citep{PhysRevD.95.062002}. The simulated signal initiates a response that mimics that
of a true GW. By looking for discrepancies between the
injected and recovered signals, it is possible to characterize the
performance of analyses and the coupling of instrumental subsystems to
the detectors' output channels.

During the third observing run O3, hardware injections were performed
in the Advanced LIGO and Advanced Virgo detectors. The record of all
injections is available through GWOSC web pages.\footnote{\url{http://gwosc.org/O3/o3_inj}}
This list is provided to prevent potential confusion with an actual
astrophysical signal.
For Virgo, those injections were removed post-facto when producing the calibrated strain (see \citet{Acernese_2022} for details on this subtraction), so the injection
times are not marked in the GWOSC files. On the other hand, in the
case of Advanced LIGO the injections are still present in the
calibrated data, and their times are marked in the GWOSC files (see
Sec.~\ref{sec:DqInj}).

No injections were performed during O3GK. 


\section{Strain data}
\label{sec:record}

All O3 open data are distributed under the Creative Commons Attribution International Public License 4.0, including strain data from O3a~\citep{gwosc-O3a}, O3b~\citep{gwosc-O3b} and O3GK~\citep{gwosc-O3GK}.  Small batches of files can be conveniently downloaded from the GWOSC website directly.\footnote{\url{https://gwosc.org/data}} However, when downloading large amounts of data (such as an entire observing run) the use of the distributed file system \texttt{CernVM-FS} \citep{CVMFSpaper} is recommended.\footnote{For \texttt{CernVM-FS} installation instructions, see \url{https://gwosc.org/cvmfs}.} Once configured, \texttt{CernVM-FS} allows access to all GWOSC data locally on the user's computer.

The O3 calibrated strain data are distributed in files that contain 4096 seconds of data. Published GW signals are also released in separate files containing 
data snippets of 4096 seconds or 32 seconds, centered on the event's detection time and released under the GWOSC Event Portal.\footnote{See \url{https://gwosc.org/eventapi}.} The description of the data records that follows is valid both for single event releases and for bulk data releases.

GWOSC calibrated strain data are repackaged from data stored in the LVK archives. The data source is uniquely identified by a channel name and a frame type (see Table~\ref{tab:source}). At times when data are unavailable or of quality too poor to be analyzed, the strain values are represented with \texttt{NaN}s. Strain data are 
made available both at the sampling rate of 16384 Hz, and at a 
downsampled rate of 4096 Hz\footnote{For simplicity, in the rest of the paper the sampling rates will be indicated in kHz and rounded to the closest integer, i.e.\ 4 and 16~kHz means 4096 and 16384~Hz, respectively.}. Down-sampling is achieved using the standard decimation method implemented in \texttt{scipy.signal.decimate}\footnote{This method applies an anti-aliasing filter based on an order-8 Chebychev type I infinite impulse response (IIR) filter \citep{IIR} before decimation.} from the Python package \SCIPY{} \citep{Virtanen:2019joe}. The highest frequency available is determined by the Nyquist--Shannon sampling theorem \citep{Nyquist1}, and is equal to half the sampling rate specified in a particular dataset. This is an important consideration to keep in mind when deciding which sample rate to download from GWOSC. Because the anti-aliasing filters used in resampling roll-off at the upper end of the working frequency interval, the valid frequency range is reduced to a bit less than the Nyquist frequency. So, for the 4 kHz data the maximum usable frequency is approximately 1700 Hz. Higher sample rate data will require more hard-drive space to store and longer times to download. The user can decide which dataset meets their needs.

\begin{table}[ht]
\centering
\caption{The channel names and frame types listed in this table are unique identifiers in the LIGO, Virgo, \GEO~and KAGRA data archives that allow tracing the provenance of the strain data released on GWOSC. H1 and L1 indicate the two LIGO detectors (Hanford and Livingston respectively), V1 refers to Virgo, G1 refers to \GEO~and K1 refers to KAGRA. The attribute \texttt{CLEAN-SUB60HZ} in H1 and L1 indicates that the noise subtraction procedure described in \citet{O3clean} was used. The attributes \texttt{C01}, \texttt{V1Online} and \texttt{V1O3Repro1A} refer to the calibration version.}
\vspace{10pt}
\label{tab:source}
\resizebox{\textwidth}{!}{\begin{tabular}{llll}
\hline
Run	&	Det.	&	Channel name					&	Frame type\\
\hline
\hline
O3a	&	H1	&	\texttt{H1:DCS-CALIB\_STRAIN\_CLEAN-SUB60HZ\_C01}	&	\texttt{H1\_HOFT\_CLEAN\_SUB60HZ\_C01}\\
O3a	&	L1	&	\texttt{L1:DCS-CALIB\_STRAIN\_CLEAN-SUB60HZ\_C01}	&	\texttt{L1\_HOFT\_CLEAN\_SUB60HZ\_C01}\\
O3a	&	V1	&	\texttt{V1:Hrec\_hoft\_16384Hz}			&	\texttt{V1Online}\\
O3a(last two weeks)	&	V1	&	\texttt{V1:Hrec\_hoft\_V1O3ARepro1A\_16384Hz}	&	\texttt{V1O3Repro1A}\\
\hline
O3b	&	H1	&	\texttt{H1:DCS-CALIB\_STRAIN\_CLEAN-SUB60HZ\_C01}	&	\texttt{H1\_HOFT\_CLEAN\_SUB60HZ\_C01}\\
O3b	&	L1	&	\texttt{L1:DCS-CALIB\_STRAIN\_CLEAN-SUB60HZ\_C01}	&	\texttt{L1\_HOFT\_CLEAN\_SUB60HZ\_C01}\\
O3b	&	V1	&	\texttt{V1:Hrec\_hoft\_16384Hz}			&	\texttt{V1Online}\\
\hline
O3GK	&	G1	&	\texttt{G1:DER\_DATA\_HD\_CLEAN}		&	\texttt{G1\_RDS\_C02\_L3}\\
O3GK	&	K1	&	\texttt{K1:DAC-STRAIN\_C20}			&	\texttt{K1\_HOFT\_C20}\\
\hline
\end{tabular}
}
\end{table}  

\subsection{GWOSC file formats}
The GWOSC open data are delivered in two different file formats:  \texttt{hdf} and  \texttt{gwf}. The Hierarchical Data Format \texttt{hdf} \citep{hdf5} is a portable data format readable by many programming languages. 
The Frame format \texttt{gwf} \citep{frame} is a specialized format used by the gravitational wave community. 
Data associated with GW events are also released as plain text files containing two columns with the global positioning system (GPS) time in the first column and the corresponding strain value in  the second column. 

For both formats the file naming follows the naming convention, 
\begin{center}
\textit{obs}---\textit{FrameType}---\textit{GPSstart}---\textit{duration}.\textit{extension}
\end{center}
where \textit{FrameType} for the main O3 data release is
\begin{center}
\textit{ifo}\_\texttt{GWOSC}\_\textit{ObservationRun}\_\textit{s}\texttt{KHZ}\_\texttt{R}\textit{n}
\end{center}
and 
\begin{itemize}
\item \textit{obs} is the observatory, i.e. the site, so can have values L, H, V, G or K; 
\item \textit{ifo} is the interferometer and can have values H1, L1, V1, G1 or K1; 
\item \textit{ObservationRun} encodes the observing run name, so in this case is O3a, O3b, or O3GK; 
\item \textit{s} is the sampling rate in kHz with either a value 4 or 16 (4096~Hz or 16384~Hz); 
\item \textit{n} is the version number of the file (typically 1); 
\item \textit{GPSstart} is the starting time of the data contained in the file, as a 10-digit GPS value (in seconds); 
\item \textit{duration} is the duration in seconds of the file, typically either 4096 or 32 seconds; 
\item and \textit{extension} 
represents the file format and can be \texttt{gwf} or \texttt{hdf}.
\end{itemize}

The folders (or groups) included in the \texttt{hdf} files are:
\begin{itemize}
\item \emph{meta}: metadata of the file containing the following fields: 
	\begin{itemize}
		\item \emph{Description}, e.g. ``Strain data time series from LIGO'',
		\item \emph{DescriptionURL}: URL of the GWOSC website,
		\item \emph{Detector}, e.g. L1, and \emph{Observatory}, e.g. L,\footnote{The observatory refers to the site and it is indicated by one letter, like L for Livingston. The addition of a number after the letter to indicate the detector, e.g. L1, could be useful if multiple detectors are installed in the same site, as was the case for Initial LIGO \citep{Abbott_2009}.}
		\item \emph{Duration}, \emph{GPSstart}, \emph{UTCstart}: duration and starting time (using GPS and UTC standards, respectively) of the segment of data contained in a file.
		\item \emph{StrainChannel}: channel name used in the LVK archives
		\item \emph{FrameType}: frame type used in the LVK archives
		
	\end{itemize}

\item \emph{strain}: array of $h(t)$, sampled at 4 or 16 kHz depending on the file. For the times when the detector is not in science mode or the data does not meet the minimum required data quality conditions (see next section), the strain values are set to \texttt{NaN}s. The strain $h(t)$ is a function of time, so it is accompanied by the attributes \emph{Xstart} and \emph{Xspacing} defining the starting GPS time of the data contained in the array and the corresponding distance in time between the points of the array.

\item \emph{quality}: this folder contains two sub-folders, one for data quality and the other for injections, each including a bitmask to indicate at each second the status of the data quality or the injections and the description of each bit of the mask (see Section \ref{sec:DqInj} for details).
  
\end{itemize}

The \texttt{gwf} files have a similar content but with a different structure. They contain 3 channels, one for the strain data, one for the data quality and one for the injections. The channel names are described in Table~\ref{tab:channels}. The original files produced internally, whose channel names are listed in Table~\ref{tab:source}, contain only the strain channel, while the GWOSC files conveniently combine the strain data with the data quality and injection information in the same file. 

\begin{table}[ht]
\centering
\caption{Channel names of the GWOSC frame files (format \texttt{gwf}). In this nomenclature, \textit{ifo} is a place holder for the interferometer name, i.e. H1, L1, V1, G1 or K1, and $s$ = 4 or 16 kHz denotes the sampling rate. The \texttt{R1} sub-string represents the revision number of the channel name so it will become \texttt{R2} in case there is a second (revised) release, and so on.}
\vspace{10pt}
\label{tab:channels}
\begin{tabular}{ll}
\hline
			&	Channel name\\
\hline
\hline
Strain                  &       \textit{ifo}\texttt{:GWOSC-}$s$\texttt{KHZ\_R1\_STRAIN}\\
Data quality mask       &       \textit{ifo}\texttt{:GWOSC-}$s$\texttt{KHZ\_R1\_DQMASK}\\
Injections mask         &       \textit{ifo}\texttt{:GWOSC-}\texttt{$s$KHZ\_R1\_INJMASK}\\
\hline
\end{tabular}
\end{table}  

\subsection{Data quality and injections in GWOSC files}
\label{sec:DqInj}
Several types of searches are performed on the LIGO, Virgo, \GEO, and KAGRA data. Those searches are divided into four families named after the types of signals they target: Compact binary coalescences (\texttt{CBC}), GW bursts (\texttt{BURST}), continuous waves (\texttt{CW}) and stochastic backgrounds (\texttt{STOCH}). As each type of search has a unique sensitivity to instrumental artifacts, a detailed characterization of detector noise and data quality is essential to eliminate spurious signals of terrestrial origin found by the searches. LIGO, Virgo, \GEO~and KAGRA have dedicated teams responsible for detector characterization and data quality, as described in \citet{LIGO:detcharO2O3, Virgo:2022fxr, kagra-GEO600-2022}, and \citet{KAGRAnoise}.

\texttt{CBC} analyses \citep{2021PhRvX..11b1053A,LIGOScientific:2021usb,2021arXiv211103606T} 
seek signals from merging neutron stars and black holes by filtering the data with waveform templates. 
\texttt{BURST} analyses \citep{KAGRA:2021tnv,KAGRA:2021bhs} 
search for generic GW transients with minimal assumptions on the source or signal morphology by identifying excess power in the time-frequency representation of the GW strain data.
\texttt{CW} searches \citep{LIGOScientific:2022pjk} look for long-duration, continuous, periodic GW signals from asymmetries of rapidly spinning neutron stars. 
\texttt{STOCH} searches \citep{LIGOScientific:2021sto1,LIGOScientific:2021sto2} target the stochastic GW background signal which is formed by the superposition of and unresolved sources from various stages of the evolution of the universe.

Because of fundamental differences in the search methodologies, certain noise types are relevant to specific searches.  \texttt{CBC} and \texttt{BURST} searches
 look for short, transient signals, with durations from less than a second to several tens of seconds. 
Data quality information for these searches is recorded as sets of time intervals
when data are relatively free of corruption, known as segment lists. 
This information is provided inside the GWOSC files for the two GW transient searches \texttt{CBC} and \texttt{BURST}.  The data quality information most relevant for \texttt{CW} and \texttt{STOCH} searches is in the frequency domain and it is provided as lists of instrumental lines in separate files, 
available for download on GWOSC\footnote{See \url{https://gwosc.org/O3/o3speclines} for L1, H1 and V1, \url{https://gwosc.org/O3/O3GKspeclines} for K1 and \url{https://gwosc.org/O3/O3GK_GEO_speclines} for G1.}.

Data quality and signal injection information for a given GPS second is indicated by bitmasks with a 1-Hz sampling rate. The bit meanings are given in Tables~\ref{tab:dq} and \ref{tab:inj} for the data quality and injections, respectively. To describe data quality, different \emph{categories} are defined. For each category, the corresponding bit in the bitmask shown in Table~\ref{tab:dq} has 
a value of 1 (good data) if in that second of time the requirements of the category are fulfilled, otherwise 0 (bad data). 

\begin{table}[ht]
\centering
\caption{Data quality bitmasks description. For O3, the \texttt{CBC\_CAT1} and \texttt{BURST\_CAT1} segment lists are equivalent (see the definition of \texttt{CAT1} in the text). Note that any data that are not present are replaced by \texttt{NaN} values in the corresponding strain time series.  In each bit mask, a value of 1 corresponds to the data quality check passing (good data), and a zero means the check has failed (bad data).  The \texttt{CBC\_CAT3} and \texttt{BURST\_CAT3} are equivalent to \texttt{CBC\_CAT2} and \texttt{BURST\_CAT2} in O3.}
\vspace{10pt}
\label{tab:dq}
\begin{tabular}{lll}
\hline
Bit	&	Short name	&	Description	\\
\hline
\hline
0	& 	\texttt{DATA}		&	Data present	\\
1 	&	\texttt{CBC\_CAT1} 	&	Pass \texttt{CAT1} test\\
2 	&	\texttt{CBC\_CAT2} 	&	Pass \texttt{CAT1} and \texttt{CAT2} test for \texttt{CBC} searches\\
3 	&	\texttt{CBC\_CAT3} 	&	Pass \texttt{CAT1} and \texttt{CAT2} and  \texttt{CAT3} test for \texttt{CBC} searches\\
4 	&	\texttt{BURST\_CAT1} 	&	Pass \texttt{CAT1} test\\
5 	&	\texttt{BURST\_CAT2} 	&	Pass \texttt{CAT1} and \texttt{CAT2} test for \texttt{BURST} searches\\
6 	&	\texttt{BURST\_CAT3} 	&	Pass \texttt{CAT1} and \texttt{CAT2} and  \texttt{CAT3} test for \texttt{BURST} searches\\
\hline
\end{tabular}
\end{table}  

\begin{table}[ht]
\centering
\caption{Meaning of the injection bits.  A value of 1 indicates \texttt{TRUE} (no injection), while a value of 0 is \texttt{FALSE} (injection is present).}
\vspace{10pt}
\label{tab:inj}
\begin{tabular}{lll}
\hline
Bit	&	Short name				&	Description	\\
\hline
\hline
0	& 	\texttt{NO\_CBC\_HW\_INJ}		&	No \texttt{CBC} injections	\\
1 	&	\texttt{NO\_BURST\_HW\_INJ} 		&	No burst injections\\
2 	&	\texttt{NO\_DETCHAR\_HW\_INJ} 		&	No detector characterization injections\\
3 	&	\texttt{NO\_CW\_HW\_INJ}  		&	No continuous wave injections\\
4 	&	\texttt{NO\_STOCH\_HW\_INJ} 		&	No stochastic injections\\

\hline
\end{tabular}
\end{table}  

The meaning of each category is described in \citet{LIGO:detcharO2O3} and \citet{Virgo:2022fxr}.  Here, we provide a brief summary of each category:

\begin{description}
\item[\texttt{DATA}] Failing this level indicates that strain data are not 
publicly available at this time because the instruments were not operating in nominal conditions. For O3, this is equivalent to failing Category 1 criteria, defined below. 
For intervals of bad or absent data, \texttt{NaN}s have been inserted in the 
corresponding strain data array. 
\item[\texttt{CAT1}](Category 1) Failing a data quality check at this category indicates a critical issue with a key detector component not operating in its nominal configuration.  GWOSC data during times that fail \texttt{CAT1} criteria are replaced by \texttt{NaN} values in the strain time series.
For O3, \texttt{CBC\_CAT1}, \texttt{BURST\_CAT1}, and 
\texttt{DATA} lead to identical segment lists. 
\item[\texttt{CAT2}](Category 2) Failing a data quality check at this category indicates times when excess noise is present in a sensor with an understood physical coupling to the strain channel \citep{catO1}.  The fraction of time removed by this category is less than 
1\% of the data, and is detailed in Table \ref{table:cat2}.

\item[\texttt{CAT3}] (Category 3) Failing a data quality check at this category indicates times when there is statistical coupling between a sensor/auxiliary channel and the strain channel which is not fully understood. This category was not used in O3 LVK searches, although it was used in previous observing runs \citep{RICHABBOTT2021100658}.
\end{description}

\begin{table}[ht]
\centering
\caption{Total time satisfying the data quality criteria for each \texttt{SEARCH} type (= \texttt{CBC} or \texttt{BURST}) and each \texttt{CATEGORY} (= \texttt{CAT1}, \texttt{CAT2} or \texttt{CAT3}) spanning the full \texttt{DURATION} of each observing \texttt{RUN} (= O3a, O3b or O3GK) and each  \texttt{DETECTOR} (= H1, L1, V1, G1 or K1). \texttt{DURATION} includes all time in seconds between the official start and end of each \texttt{RUN}, including times when the instruments are not collecting data for astrophysical analysis.  When the criteria for a given flag is satisfied, the corresponding bit will have the value 1 (good data by these criteria); otherwise, it will have the value 0 (bad data). The data in the table can be retrieved at \footnotesize{\url{https://gwosc.org/timeline/show/[RUN]_16KHZ_R1/[DETECTOR]_[SEARCH]_[CATEGORY]}}.}
\vspace{20pt}
\label{tab:dqflagtime}
\resizebox{\textwidth}{!}{\begin{tabular}{cccccccccc}
\hline
\multicolumn{10}{c}{\bf Data Quality Flags (Total time in seconds)} \\
\hline
\hline
\texttt{RUN} & \texttt{DURATION} & \texttt{DETECTOR} & \texttt{DATA} & \texttt{CBC\_CAT1} & \texttt{CBC\_CAT2} & \texttt{CBC\_CAT3} & \texttt{BURST\_CAT1} & \texttt{BURST\_CAT2} & \texttt{BURST\_CAT3} \\
\hline
    &          & H1 & 11218675 & 11218675 & 11177046 & 11177046 & 11218675 & 11125849 & 11125849 \\
{\bf O3a} & 15811200 & L1 & 11956179 & 11956179 & 11943913 & 11943913 & 11956179 & 11879365 & 11879365 \\
    &          & V1 & 12038929 & 12038929 & 12038929 & 12038929 & 12038929 & 12038929 & 12038929 \\
\hline
    &          & H1 & 9967195 & 9967195 & 9964945 & 9964945 & 9967195 & 9915276 & 9915276 \\
{\bf O3b} & 12708000 & L1 & 9810816 & 9810816 & 9782946 & 9782946 & 9810816 & 9760960 & 9760960 \\
    &          & V1 & 9591207 & 9591207 & 9591207 & 9591207 & 9591207 & 9591207 & 9591207 \\
\hline
\multirow{2}{*}{\bf O3GK} & \multirow{2}{*}{1180800} & G1 & 940133 & 940133 & 940133 & 940133 & 940133 & 940133 & 940133 \\
                      &                          & K1 & 628055 & 628055 & 628055 & 628055 & 628055 & 628055 & 628055 \\
\hline
\end{tabular}
}
\end{table}

\begin{table}[h!]
  \caption{Fraction of observing time removed by applying \texttt{CAT2} vetoes.  The percentages represent the amount of time in the \texttt{DATA} segment list relative to the total duration of observing time. \texttt{CAT2} vetoes were not used for Virgo, KAGRA, or \GEO.}
  \label{table:cat2}
  \centering
\begin{tabular}{c c c c c } 
\hline
Detector & \multicolumn{2}{c}{O3a} & \multicolumn{2}{c}{O3b} \\
& \texttt{CBC} & \texttt{BURST} & \texttt{CBC} & \texttt{BURST}\\ 
 \hline\hline
 H1 & 0.37\% & 0.83\% & 0.02\% & 0.52\% \\ 
 \hline
 L1 & 0.01\% & 0.64\% & 0.28\% & 0.51\%  \\
  \hline
\end{tabular}
\end{table}

Data quality categories are cascading: a time which fails a given category automatically fails all higher categories. Since \texttt{CAT3} flags were not used in O3, the \texttt{CAT3} segment lists are identical to the
corresponding \texttt{CAT2} lists.  However, the different analysis groups qualify the data independently: failing \texttt{BURST\_CAT2} does not necessarily imply failing \texttt{CBC\_CAT2}. See Table~\ref{tab:dqflagtime} for the
amount of time associated with each category.

Simulated signals added to the detectors for testing and calibration  are referred to as \textit{hardware injections}. GWOSC data releases
provide a time series with each one second sample representing a bit mask vector of the state of the injection at that time. The injections
are categorized according to the type of injected signal relevant to each astrophysical search. There are also injections used for detector characterization
(\texttt{DETCHAR}). The injection bitmask marks the injection-free times. The bit corresponding to a given type of injection is defined in
Table~\ref{tab:inj}. A bit is set to 1 if there is no injection, otherwise it is set to 0. The full details of the complete set of hardware injections for O3 can be found at \url{https://gwosc.org/O3/o3_inj}.

A \texttt{CBC} signal was injected on one occasion in each of the LIGO detectors during O3a, but not in coincidence. The 
(\texttt{CBC}) injection for H1 took place between GPS 1251662270 (2019-09-04T19:57:32 UTC) and GPS 1251662279 (2019-09-04T19:57:41 UTC), which 
was during observing mode. The \texttt{CBC} injection for L1 did not take place during observing mode and is thus not present in the released data. No \texttt{CBC} injections were performed during O3b. 

No injections of \texttt{BURST} signals were performed during O3a and O3b.

Some signal injections of the \texttt{DETCHAR} type were injected during O3a and O3b for both H1 and L1, while there were none for Virgo.
The waveform model used for those injections is $h(t) = a\ e^{-(t-t_0)^2 / {\tau}^2} \sin{[2 \pi f (t-t_0) - \phi]}$, where $t_0$ is the time of the injection.
The signal parameters $\tau$, $\phi$ and $f$ were randomly chosen and are documented in separate databases for H1\footnote{See \url{https://gwosc.org/static/injections/o3a/H1_detchar_inj.txt} for O3a and \url{https://gwosc.org/static/injections/O3b/inj_o3b_H1.txt} for O3b.}
and for L1\footnote{See \url{https://gwosc.org/static/injections/o3a/L1_detchar_inj.txt} for O3a and \url{https://gwosc.org/static/injections/O3b/inj_o3b_L1.txt} for O3b.}.

Four \texttt{STOCH} signals were injected during O3. These hardware injections consisted of a simulated stochastic gravitational-wave background of 13 minutes long duration and were generated using the NAP package \citep{NAPtool} and rescaled to have an amplitude of ${\Omega}_0 = 2.0 \times 10^{-5}$ with the default Hubble constant value $H_0 = 100\ \mathrm{km}\ \mathrm{s}^{-1}\ \mathrm{Mpc}^{-1}$. Two of these injections were added only in H1 starting at GPS 1249200018 (2019-08-07T08:00:00 UTC)  and GPS 1258273818 (2019-11-20T08:30:00). The other two injections were performed coherently at L1 and H1 during O3b starting at GPS 1258345818 (2019-11-21T04:30:00) and GPS 1258353018 (2019-11-21T06:30:00). The detectors were in observing mode for all of these stochastic injections.

\texttt{CW} injections were performed in H1 and L1 during both O3a and O3b, using a set of pulsar parameters provided on the GWOSC website\footnote{\url{https://gwosc.org/O3/O3April1_injection_parameters}}. Those injections are always present except during
defined intervals for O3a\footnote{\url{https://gwosc.org/timeline/show/O3a_16KHZ_R1/H1_NO_CW_HW_INJ*H1_DATA*L1_NO_CW_HW_INJ*L1_DATA/1238166018/15811200}}
and O3b\footnote{\url{https://gwosc.org/timeline/show/O3b_16KHZ_R1/H1_NO_CW_HW_INJ*H1_DATA*L1_NO_CW_HW_INJ*L1_DATA/1256655618/12708000}}.
No \texttt{CW} injections were performed in V1 during O3a, but there were \texttt{CW} injections in V1 during O3b between GPS 1263945616 (2020-01-24 23:59:58 UTC) and GPS 1266019220 (2020-02-18 00:00:02 UTC). The injected signal was removed a posteriori in the strain data as described in \citet{Acernese_2022}.
A residual signal after removal may still be present with an amplitude between 20 and 100 times lower than that of the injection. The residual amplitude is smaller than or at most the same order as the calibration uncertainty.

No signal hardware injections of any type were performed during O3GK.

\subsection{Alternate versions of the strain data}
\label{sec:AlternateStrain}

\begin{table}[ht]
\centering
\caption{Names of alternate strain channels in the O3 data release. \textit{ifo} is a place holder for the name of the LIGO interferometer, i.e. H1 or L1.}
\vspace{10pt}
\label{tab:altcal}
\begin{tabular}{lp{8cm}}
\hline
Channel name &	Description\\
\hline
\hline
\textit{ifo}\texttt{:DCS-CALIB\_STRAIN\_C01\_AR} & LIGO calibrated strain, offline calibration\\
\textit{ifo}\texttt{:DCS-CALIB\_STRAIN\_CLEAN\_C01\_AR} & LIGO calibrated strain, after applying linear noise subtraction\\
\textit{ifo}\texttt{:DCS-CALIB\_STRAIN\_CLEAN\_SUB60HZ\_C01\_AR} & LIGO calibrated strain, after applying both linear and non-linear noise subtraction. 
This is the recommended channel (main release)\\
\texttt{V1:Hrec\_hoft\_16384Hz\_AR} & Virgo calibrated strain for most of O3a and O3b (main release)\\
\texttt{V1:Hrec\_hoft\_V1O3ARepro1A\_16384Hz\_AR} & Virgo calibrated strain for the last two weeks of Sep 2019, near the end of O3a with an enhanced 
noise subtraction (main release)\\
\hline
\end{tabular}
\end{table}

In addition to the main strain data release described above, the O3 data release includes several alternate strain channels, as described 
at \url{https://gwosc.org/O3/O3alt}.  This alternate data release is available 
via both \verb=CernVM-FS= or streaming via a network data server (NDS2) \citep{nds2_wiki_page}.
The alternate strain channels reflect different choices for how aggressively to apply noise subtraction strategies to remove different
sources of contamination.  Some LVK analyses used different versions of the strain 
channels.  The alternate strain channel release was designed to reflect the internal formatting 
used by the LVK as much as possible. In particular, the release uses only the \verb=GWF= 
file format, does not include any \verb=NaN= values, and does not include any data quality 
information.  The channels found in the alternate calibration release are described in Table~\ref{tab:altcal}.

\section{Online event catalogs}
\label{sec:eventportal}

Ninety-three GW transient events or notable candidates were discovered based on the LVK's analyses of the O3 data
\citep{2021PhRvX..11b1053A,LIGOScientific:2021usb,2021arXiv211103606T}.
Data associated with these signals are available online through the GWOSC Event Portal\footnote{\url{https://gwosc.org/eventapi}}, 
along with other scientific products.  
For all events in the Event Portal, snippets of strain data are
released in the form of a segment of 4096 seconds around the time of the
event. The data snippets are made available no later than when the event
discovery becomes public in a refereed, scientific journal.
In addition,
the Event Portal includes a concise summary of the source properties (i.e., parameters
of the compact star binaries associated with each of the detected
signals), links to a
number of science products (posterior samples), links to any associated low-latency alerts,
and a documentation page for each release containing publication information.
The list of O3 event data releases is as follows:
\begin{description}

\item[\texttt{O3\_Discovery\_Papers}] Notable events first published individually \citep{LIGOScientific:2020aai,LIGOScientific:2020stg,LIGOScientific:2020iuh,LIGOScientific:2020zkf,LIGOScientific:2021qlt}. Associated data releases may contain preliminary versions of data quality segments and calibration

\item[\texttt{O3\_IMBH\_marginal}] Marginal candidates associated with the search for Intermediate Mass Black Hole (IMBH) binary mergers

\item[\texttt{GWTC-2}] Confident events from the O3a observation run (first search)

\item[\texttt{GWTC-2.1-confident}] Confident events from the O3a observation run (updated search)

\item[\texttt{GWTC-2.1-marginal}] Marginal candidates from the O3a observation run (updated search)

\item[\texttt{GWTC-2.1-auxiliary}] Candidates from GWTC-2 which, based on the updated analysis presented in the GWTC-2.1 catalog paper, do not satisfy the criteria for inclusion in the GWTC-2.1-confident or GWTC-2.1-marginal releases

\item[\texttt{GWTC-3-confident}] Confident events from the O3b observing run

\item[\texttt{GWTC-3-marginal}] Marginal candidates from the O3b observing run 

\end{description}

Some events are listed in the database with multiple versions, typically corresponding to the event's inclusion in multiple releases.
The cumulative \texttt{GWTC} catalog includes all confident GW events published by the LVK collaboration, and currently includes 93 events.  Events in the \texttt{GWTC-2.1-confident} and \texttt{GWTC-3-confident} releases all have a probability of astrophysical origin greater than 0.5 \footnote{See \citet{2021arXiv211103606T}, App. D.~7 for a definition and details about its estimation.}
in at least one of the search pipelines, and are included in the cumulative \texttt{GWTC}.

The online catalogs are searchable via a web user interface.  
The Event Portal database can be queried based on specific source properties, 
namely the primary mass, secondary mass, total mass, chirp mass, final
mass (of the merger remnant), luminosity distance, redshift, effective
inspiral spin, or other properties associated with the observed signal, such as
UTC or GPS event time, detector frame chirp mass, network SNR, false
alarm rate and the posterior probability of astrophysical origin.
The events can also be selected by identification such as
partial event name, release catalog or group of catalogs.  The output
format can be one of the following: HTML, JSON, CSV or plain ASCII
text.

To ease the analysis of multiple events, the catalogs can be queried
programmatically with scripts using the REST API that returns all
catalog lists in a JSON format.  Catalogs can be queried with a
\texttt{GET} request.  As an example, to request all merger events for
which the primary mass is less than $3 M_{\odot}$, the URL for the
\texttt{GET} request would be
\href{https://gwosc.org/eventapi/html/query/show?max-mass-1-source=3}{https://gwosc.org/eventapi/html/query/show?max-mass-1-source=3}.
A detailed explanation of the query API nodes can be found on the
GWOSC website\footnote{\url{https://gwosc.org/apidocs}}.

\subsection{Parameter estimation}

For each detected source the Event Portal displays the 90\% credible intervals for a selection of parameters that reflect the values given in the relevant publication. Those credible intervals are computed from the posterior samples resulting from Bayesian inference algorithms applied to the data.

In addition to the information provided by GWOSC, the posterior samples are
distributed through the Zenodo open repository \citep{zenodo_2021_5546665, zenodo_2021_5759108}.

They are provided from the single event web page and through the JSON API as downloadable links to the files on Zenodo. The parameter names follow a standard nomenclature\footnote{See \url{https://lscsoft.docs.ligo.org/pesummary/unstable_docs/gw/parameters.html}.} \citep{Hoy:2020vys}. Parameter estimates may change with different version of the event or catalog release.  The parameter sets are denoted by a set of version numbers for each event (depending on the number of releases in which that event appears).

\subsection{Low-latency alerts}

During O3, public alerts 
were communicated with low latency to report
the occurence of a notable trigger detected in the data\footnote{See \url{https://emfollow.docs.ligo.org/userguide} for more details. This userguide is a living document that is being updated in preparation for the upcoming science run O4. Therefore, the informations in this guide may not be necessarily relevant for O3 data.}. The alerts
are sent with a latency of few minutes after detection. They include a
number of preliminary parameter estimations that are useful for the
localization of the source through a probability skymap. This
information can be used by other, non-GW, instruments to search for potential
electromagnetic counterparts in follow-up observations.  The complete
list of alerts sent during O3 can be found publicly in the GraceDB
website\footnote{\url{https://gracedb.ligo.org/superevents/public/O3}}
and, as described below, in GWOSC.

The Event Portal references the GraceDB entry for the original
trigger alert of the event. Links to GraceDB entries are available
through the GWOSC web interface and the JSON API. Events first detected
offline do not trigger low-latency alerts and thus lack a GraceDB
entry.


\section{Technical Validation}
\label{sec:validation}

The O3 GWOSC data release is reprocessed for the broader user community beginning with the internal strain data products used for data analysis 
by the LIGO, Virgo, and KAGRA Collaborations for publication purposes. The reprocessing produces new GWOSC \texttt{gwf} and \texttt{hdf5} files 
containing the previously discussed strain, data quality and hardware injection information for each detector. In addition, versions of these 
GWOSC files at a reduced sampling rate of 4096 Hz for the strain channel of each detector are also produced. All data for the release are
carefully reviewed by the internal GWOSC team and then reviewed by an independent review team made up of members from the LIGO, 
Virgo, and KAGRA Collaborations. This review process checks that:
\begin{itemize}
\item the strain vectors at the maximum sample rate (16 kHz) in the GWOSC \texttt{hdf} and \texttt{gwf} files are identical to machine precision to the 
corresponding strain vectors of the LVK main archives;
\item the strain vectors after resampling at 4 kHz  do not have numerical artifacts that may arise from the resampling technique;
\item the data quality and injection information located in either the GWOSC \texttt{hdf} and \texttt{gwf} files or the online \textit{Timeline} 
tool described in detail in Section \ref{sec:usage}, agree with all available 
records. 
\item the documentation associated with the O3 data products found online is correct and contains comprehensive information for the broader user community.
\end{itemize}

The data files and accompanying documentation are released to the public on the GWOSC website once all checks have passed at the designated date and time agreed to by the LIGO, Virgo and KAGRA Collaborations.

\section{Usage notes}
\label{sec:usage}

\subsection{Salient features of GW data}

Working with GW data requires an awareness of the presence
of noise in the data.  An overview of LIGO/Virgo detector noise
and some applicable signal processing methods are described in 
\citet{LIGOScientific:2019hgc}; see also above in Sec.~\ref{sec:DqInj} 
and \ref{sec:detchar} for a brief introduction to 
various classes of detector noise. In addition, as mentioned previously,
the data are only valid within a fixed frequency range due to the limits of calibration 
(Sec.~\ref{sec:calibration})
as well as due to artifacts from the down-sampling process (Sec.~\ref{sec:record}).  
All of these complications need to be considered when searching for 
astrophysical signals.  

\subsection{List of observing segments}

\begin{figure}
\centering
\includegraphics[width=0.9\textwidth]{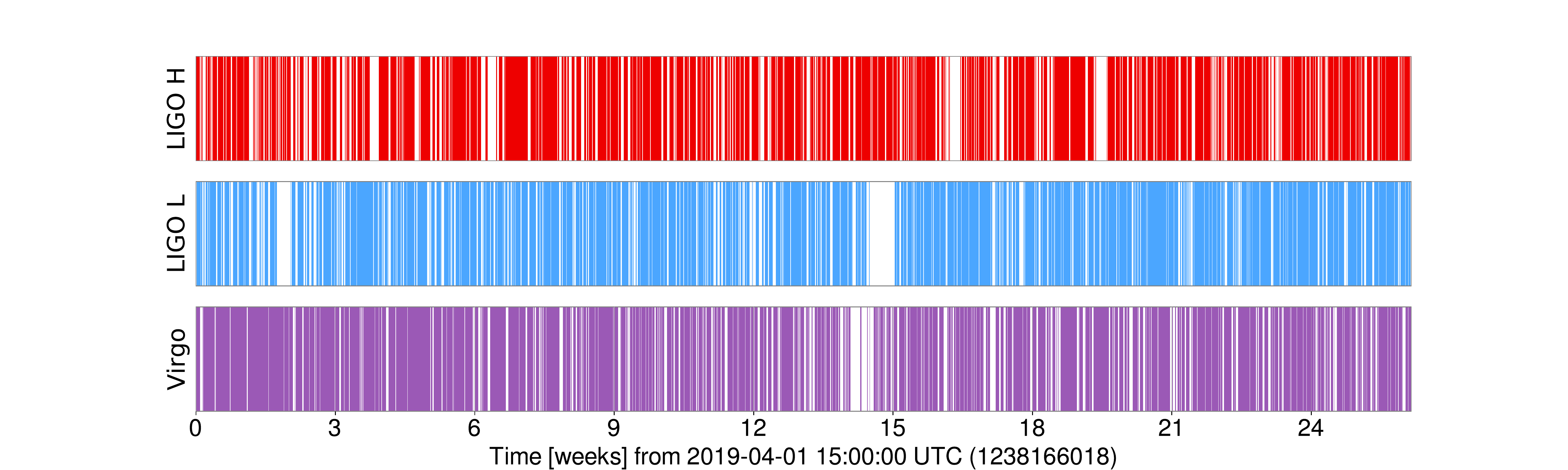}
\includegraphics[width=0.9\textwidth]{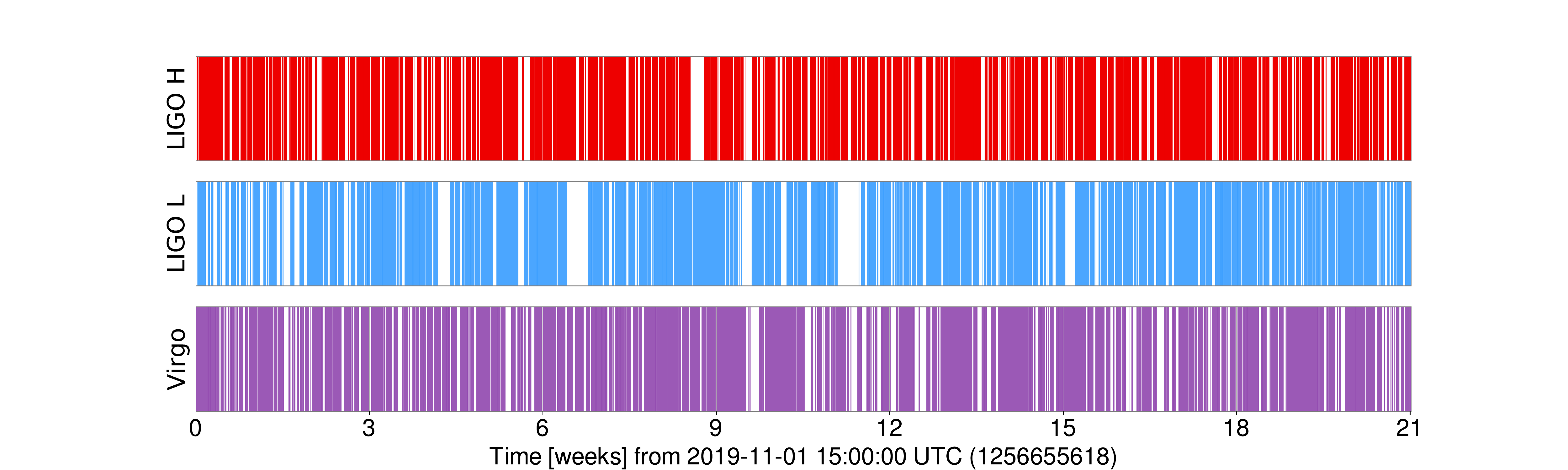}
\includegraphics[width=0.9\textwidth]{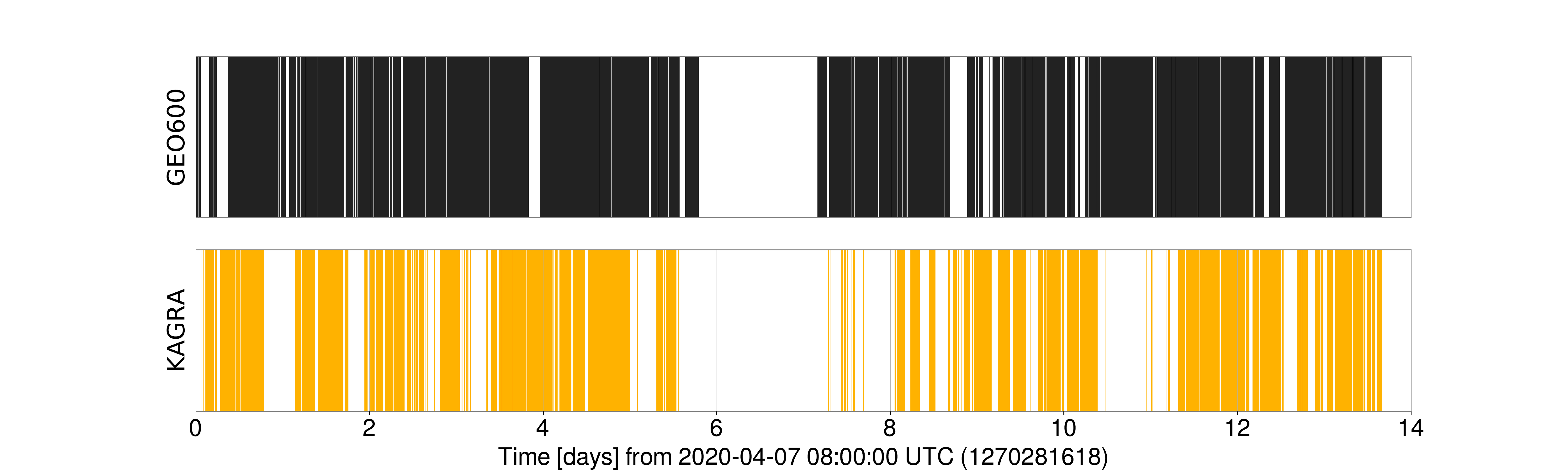}
\caption{Timelines for the full O3a (top) and O3b (middle) and O3GK (bottom) observing runs based on the data quality bitmask \texttt{CBC\_CAT1} for each detector (see Tab.~\ref{tab:dq}). Colored bars represent times when data are available, and white areas show times when data are not available. Similar plots can be generated from the GWOSC web pages\footref{fn:timeline}.}
\label{timeline-plot}
\end{figure}

Segment lists describe times when GW detectors are collecting data
and are operating in a normal condition, as described in Section \ref{sec:DqInj}.
The GWOSC website provides an online app called \verb=Timeline= to discover,
plot, and download segment lists\footnote{\label{fn:timeline}See \url{https://gwosc.org/timeline}.}.  The Timeline query page allows users to select
observing runs from a drop-down menu, and then view the names of segment lists associated
with the selected run.  Segment lists may be downloaded as ASCII text files or
in a JSON format.  Alternatively, segment lists may be displayed in an interactive
plot, as seen in Fig.~\ref{timeline-plot}.  To explore times within a run,
a visitor can use the mouse to scroll and zoom on the Timeline plots.  Hovering the mouse
over a segment displays a tool-tip with the exact start and stop time, in both GPS and
UTC time.

\subsection{Software and Support}

The GWOSC website provides a number of resources for helping investigators
learn to work with GW data, including:

\begin{itemize}
  
\item{\textbf{Software libraries}\footnote{\url{https://gwosc.org/software}}: A number of software packages developed for GW analysis are open source.
  The GWOSC website provides a suggested list of packages, many of which were created
  by members of the LIGO, Virgo, and KAGRA collaborations.  Links to source code and documentation are provided for each
  package. }

\item{\textbf{Tutorials}\footnote{\url{https://gwosc.org/tutorials}}: GWOSC provides tutorials to demonstrate the basics of GW data analysis. Most tutorials are in \verb=Python=, and provided in notebooks that can be run in the cloud to avoid the necessity for the user to install software.
}

\item{\textbf{Workshops and online course}\footnote{\url{https://gwosc.org/workshops}}: Annual Open Data Workshops provide a complete course in working with GW data, including lectures, software tutorials, and challenge problems. Materials from past workshops are available as a free online course; students can enroll at any time.  Future workshops will be posted on the GWOSC website, and are open to any interested participants.}

\item{\textbf{Discussion forum}\footnote{\url{https://ask.igwn.org}}: A public discussion forum for GW topics provides space to ask for help with GW data analysis, discuss LVK papers, post questions about GW science, and connect with other researchers in the
  field.}

\end{itemize}

\section{Summary}

The O3 data set described in this paper represents 
the most sensitive gravitational-wave observations to date.
The data contain over 80 compact object merger signals, as described in 
a number of catalog releases, including GWTC-2 \citep{2021PhRvX..11b1053A}, GWTC-2.1 \citep{LIGOScientific:2021usb} and GWTC-3 \citep{2021arXiv211103606T}.
O3 includes three main phases: O3a, O3b, and O3GK.
O3a and O3b are both joint runs of LIGO and Virgo,
while the O3GK run involved KAGRA and 
\GEO.  Data and documentation for all O3 data are
available from the GWOSC website.

Looking ahead, LIGO, Virgo, and KAGRA are planning an O4 run,
scheduled to begin in 2023, with improved sensitivity. 
Data from events discovered in O4 will be released as the events are published, and
release of the next large strain data sets are planned for 2025 \citep{dmp}.
This will be followed by the O5 observing run, anticipated to be the
first extended observing run with a span of over two years
\citep{LVK-obsplans}. Planned instrument upgrades should
increase the sensitivity of the network and thus extend the volume of space over which signals may be observed, so that future data sets will include
more frequent detections and a corresponding expanded depth of science
in this rapidly evolving field.

\begin{acknowledgments}

\section*{Acknowledgements}
    
Calibration of the LIGO strain data was performed with \GSTLAL{}-based calibration software pipeline~\citep{Viets:2017yvy}.
Calibration of the Virgo strain data was performed with C-based software~\citep{Acernese_2022}.
Data-quality products and event-validation results were computed using the \DMT{}~\citep{DMTdocumentation}, \DQR{}~\citep{DQRdocumentation}, \DQSEGDB{}~\citep{Fisher:2020pnr}, \GWDETCHAR{}~\citep{gwdetchar-software}, \HVETO{}~\citep{Smith:2011an}, \IDQ{}~\citep{Essick:2020qpo} and \OMICRONSCAN{}~\citep{Robinet:2020lbf} software packages and contributing software tools.  
Analyses relied upon the \LALSUITE{} software library~\citep{lalsuite-software}.
\PESUMMARY{} was used to post-process and collate parameter-estimation results~\citep{Hoy:2020vys}.
For an exhaustive list of the softwares used for searching the GW signals and characterizing their source, see \citep{2021arXiv211103606T}. 
Plots were prepared with \PLT{}~\citep{Hunter:2007ouj}, \SEABORN{}~\citep{Waskom:2021psk} \GWSUMM{}~\citep{gwsumm-software} and \GWPY{}~\citep{gwpy-software}. 
\NUMPY{}~\citep{Harris:2020xlr} and \SCIPY{}~\citep{Virtanen:2019joe} were used in the preparation of the manuscript.

\newif\ifcoreonly
\newif\ifkagra
\newif\ifheader
\coreonlytrue  \kagratrue  \headertrue
This material is based upon work supported by NSF’s
LIGO Laboratory which is a major facility
fully funded by the National Science Foundation.
The authors also gratefully acknowledge the support of
the Science and Technology Facilities Council (STFC) of the
United Kingdom, the Max-Planck-Society (MPS), and the State of
Niedersachsen/Germany for support of the construction of Advanced LIGO 
and construction and operation of the GEO\,600 detector. 
Additional support for Advanced LIGO was provided by the Australian Research Council.
The authors gratefully acknowledge the Italian Istituto Nazionale di Fisica Nucleare (INFN),  
the French Centre National de la Recherche Scientifique (CNRS) and
the Netherlands Organization for Scientific Research (NWO), 
for the construction and operation of the Virgo detector
and the creation and support  of the EGO consortium. 
The authors also gratefully acknowledge research support from these agencies as well as by 
the Council of Scientific and Industrial Research of India, 
the Department of Science and Technology, India,
the Science \& Engineering Research Board (SERB), India,
the Ministry of Human Resource Development, India,
the Spanish Agencia Estatal de Investigaci\'on (AEI),
the Spanish Ministerio de Ciencia e Innovaci\'on and Ministerio de Universidades,
the Conselleria de Fons Europeus, Universitat i Cultura and the Direcci\'o General de Pol\'{\i}tica Universitaria i Recerca del Govern de les Illes Balears,
the Conselleria d'Innovaci\'o, Universitats, Ci\`encia i Societat Digital de la Generalitat Valenciana and
the CERCA Programme Generalitat de Catalunya, Spain,
the National Science Centre of Poland and the European Union – European Regional Development Fund; Foundation for Polish Science (FNP),
the Swiss National Science Foundation (SNSF),
the Russian Foundation for Basic Research, 
the Russian Science Foundation,
the European Commission,
the European Social Funds (ESF),
the European Regional Development Funds (ERDF),
the Royal Society, 
the Scottish Funding Council, 
the Scottish Universities Physics Alliance, 
the Hungarian Scientific Research Fund (OTKA),
the French Lyon Institute of Origins (LIO),
the Belgian Fonds de la Recherche Scientifique (FRS-FNRS), 
Actions de Recherche Concertées (ARC) and
Fonds Wetenschappelijk Onderzoek – Vlaanderen (FWO), Belgium,
the Paris \^{I}le-de-France Region, 
the National Research, Development and Innovation Office Hungary (NKFIH), 
the National Research Foundation of Korea,
the Natural Science and Engineering Research Council Canada,
Canadian Foundation for Innovation (CFI),
the Brazilian Ministry of Science, Technology, and Innovations,
the International Center for Theoretical Physics South American Institute for Fundamental Research (ICTP-SAIFR), 
the Research Grants Council of Hong Kong,
the National Natural Science Foundation of China (NSFC),
the Leverhulme Trust, 
the Research Corporation, 
the Ministry of Science and Technology (MOST), Taiwan,
the United States Department of Energy,
and
the Kavli Foundation.
The authors gratefully acknowledge the support of the NSF, STFC, INFN and CNRS for provision of computational resources.

\ifcoreonly\else
{\bf For papers using O3b (and future) data, the following paragraph should be added for KAGRA.}\\
\fi
\ifkagra
This work was supported by MEXT, JSPS Leading-edge Research
Infrastructure Program, JSPS Grant-in-Aid for Specially Promoted
Research 26000005, JSPS Grant-inAid for Scientific Research on
Innovative Areas 2905: JP17H06358, JP17H06361 and JP17H06364, JSPS Core-to-Core
Program A. Advanced Research Networks, JSPS Grantin-Aid for
Scientific Research (S) 17H06133 and 20H05639 , JSPS Grant-in-Aid for
Transformative Research Areas (A) 20A203: JP20H05854, the joint research
program of the Institute for Cosmic Ray Research, University of Tokyo,
National Research Foundation (NRF), Computing Infrastructure Project of
Global Science experimental Data hub Center (GSDC) at KISTI, Korea
Astronomy and Space Science Institute (KASI), and Ministry of Science
and ICT (MSIT) in Korea, Academia Sinica (AS), AS Grid Center (ASGC) and
the National Science and Technology Council (NSTC) in Taiwan under
grants including the Rising Star Program and Science Vanguard Research
Program, Advanced Technology Center (ATC) of NAOJ, and Mechanical Engineering Center of KEK.
\fi

\ifcoreonly\else
For the purpose of open access, the authors have applied a Creative Commons Attribution (CC BY)
license to any Author Accepted Manuscript version arising.
We request that citations to this article use 'R. Abbott {\it et al.} (LIGO-Virgo-KAGRA Collaboration), ...' or similar phrasing, depending on journal convention.
\fi

\ifcoreonly\else
{\bf For certain collaboration papers, it may be appropriate to acknowledge specific analysis software.
One template to consider is that used for the GW190425 discovery paper, for which the latex can be
found here:}\\
{\small https://git.ligo.org/publications/gw190425/gw190425-discovery/-/blob/master/gw190425-discovery.tex\#L259.}
\fi

\ifcoreonly\else
{\bf For collaboration papers released after March 2020, it may be appropriate to add the following special acknowledgement, BUT the journal may reject it (PRL rejected it). If one chooses, one can include such acknowledgements in arXiv/DCC versions only in that case:}\\
\fi
\ifcoreonly\else
{\it We would like to thank all of the essential workers who put their health at risk during the COVID-19 pandemic, without whom we would not have been able to complete this work.}
\fi

\end{acknowledgments}

\bibliography{main}

\end{document}